\documentclass{article}
\usepackage{spconf,amsmath,graphicx}
\usepackage{multirow}
\usepackage{subfigure}
\title{Learning how to listen: A temporal-frequential attention model for sound event detection}
\name{Yu-Han Shen, Ke-Xin He, Wei-Qiang Zhang\thanks{The corresponding author is Wei-Qiang Zhang.}}
\address{
Department of Electronic Engineering, Tsinghua University, Beijing, China\\
\tt yhshen@hotmail.com, hekexinchn@163.com, wqzhang@tsinghua.edu.cn}
\begin{document}
\ninept
\topmargin=0mm
\maketitle
\begin{abstract}
In this paper, we propose a temporal-frequential attention model for sound event detection (SED). Our network learns how to listen with two attention models: a temporal attention model and a frequential attention model. Proposed system learns when to listen using the temporal attention model while it learns where to listen on the frequency axis using the frequential attention model. With these two models, we attempt to make our system pay more attention to important frames or segments and important frequency components for sound event detection. Our proposed method is demonstrated on the task 2 of Detection and Classification of Acoustic Scenes and Events (DCASE) 2017 Challenge and achieves competitive performance.
\end{abstract}

\section{Introduction}
\label{sec:intro}
Nowadays, sound event detection (SED), also named as acoustic event detection(AED), is considered as a popular topic in the field of acoustic signal processing.
The aim of SED is to temporally locate the onset and offset times of target sound events present in an audio recording.

The Detection and Classification of Acoustic Scenes and Events (DCASE) Challenge is an international
challenge concerning SED, and has been held for several years. In
DCASE 2017 Challenge, the theme of task 2 is ``detection of rare sound events" \cite{task}. It provides dataset \cite{dataset} and baseline for rare sound event detection in synthesized recordings. Here, ``rare'' means that target sound
events (babycry, glassbreak, gunshot) would occur at most once within a 30-second recording.
And the mean duration of target sound event is very short: 2.25 s for babycry, 1.16 s for glassbreak, 1.32 s for gunshot, leading to a serious problem of data imbalance.  All audio recordings are notated with ground-truth labels of event class, onset and offset time. According to the
task description, a separate system should be developed for each of the three target event classes to detect
the temporal occurrences of these events \cite{task}.

Among the submissions in DCASE 2017, most models are based on deep neural networks. Both of the top 2 teams \cite{Lim,Cakir} utilized Convolutional Recurrent Neural Networks (CRNN) as their main architecture.
They combined Convolutional Neural Networks (CNN) with Recurrent Neural Networks (RNN) to make frame-level
predictions for target events and then adopted post-processing to get the onset and offset time of sound events.
Kao et al. \cite{Kao} proposed a Region-based Convolutional Recurrent Neural Network (R-CRNN) to
improve previous work in 2018. In our work, we followed the main architecture of those three models and used CRNN as main classifier.

Inspired by the excellent performance of attention model in machine translation \cite{mt}, image caption \cite{caption}, speaker verification \cite{speaker}, audio tagging \cite{Kong}, we proposed an attention model for SED. Currently, most attention models in speech and audio processing only concentrate on time domain. We proposed a temporal-frequential attention model to focus on important frequency components as well as important frames or segments. Our attention model can learn how to listen by extracting not only temporal information but also spectral information. Besides, we visualized the weights of attention models to show what our models have actually learnt.


%
%

The rest of this paper is organized as follows: in Section 2, we introduce our methods in detail, mainly including feature extraction, baseline and temporal-frequential attention model. The dataset, experiment setup and evaluation metric are
illustrated in Section 3. The results and analysis are presented in Section 4. Finally, we conclude our work in Section 5.

\section{Methods}
\label{sec:methods}
\subsection{System overview}
\label{ssec:overview}
As shown in Figure 1, our proposed system is a CRNN architecture with temporal-frequential attention model.
 The input of our system is a 2-dim acoustic feature. It is fed into a frequential attention
 model to produce frequential attention weights. Our system learns to focus on specific frequency components of audios using those attention weights. The input acoustic feature will
 multiply with those attention weights and then pass through CRNN architecture. Compared with
 traditional CRNN \cite{Lim,Cakir}, we add a temporal attention model to let our system pay different attention
 to different frames. The temporal attention weights will multiply with the outputs of CRNN by element-wise. A sigmoid activation is used to get normalized probabilities. Then we utilize post processing to get final detection outputs.
\begin{figure}[tb]
\centerline{\includegraphics[width=0.95\linewidth]{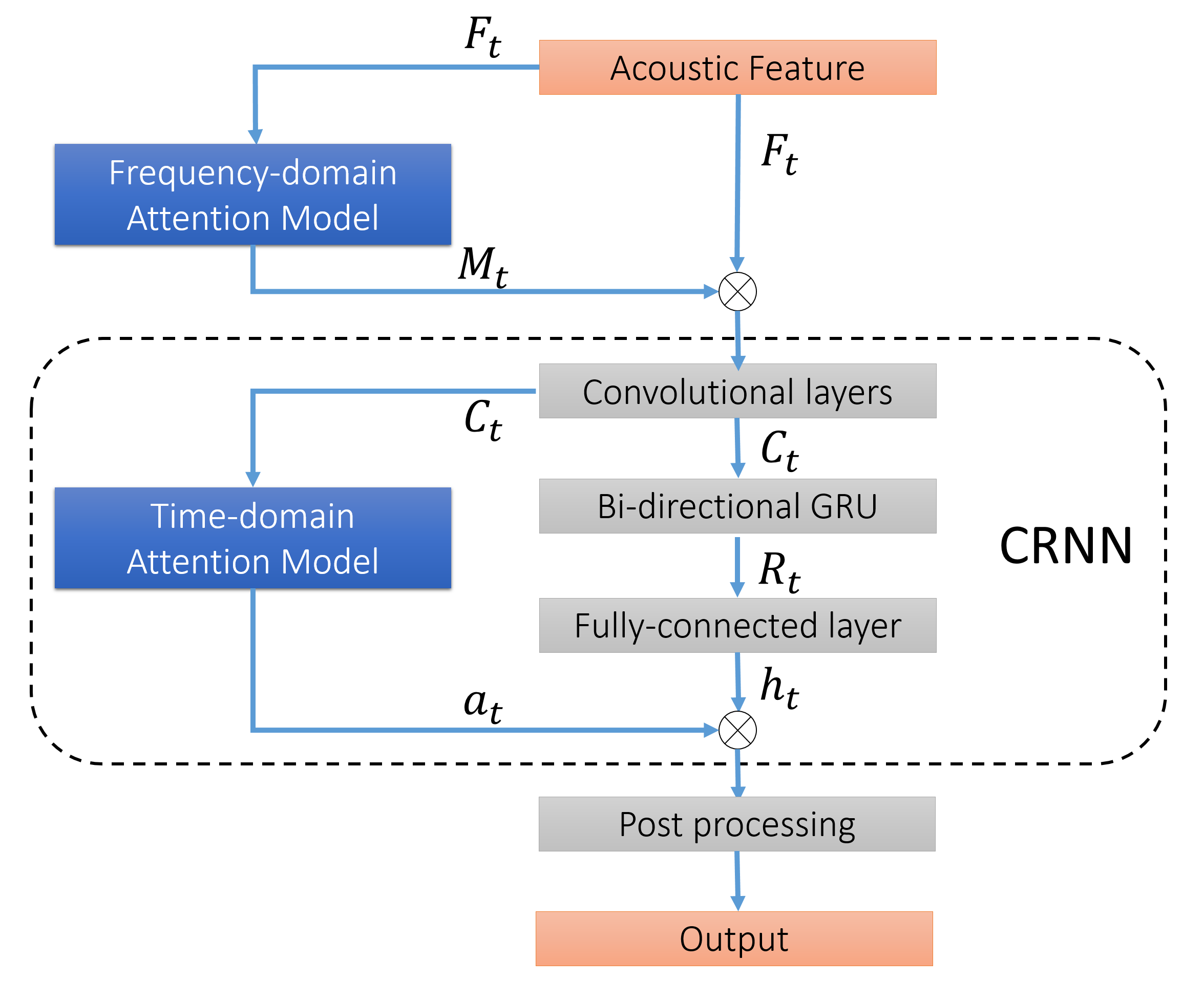}}
\caption{Illustration of overall system.}
\label{fig:fig1}
\end{figure}

\subsection{Feature extraction}
\label{ssec:feature}
The acoustic feature used in our work is log filter bank energy (Fbank). The sampling rate of input audios is 44.1kHz. To extract Fbank feature, each audio is divided into frames of 40 ms duration with shifts of 20 ms. Then we apply 128 mel-scale filters covering the frequency range 300 to 22050 Hz on the magnitude spectrum of each frame. Finally, we take logarithm on the amplitude and get Fbank feature. The extracted Fbank feature is normalized to zero mean and unit standard deviation before being fed into neural networks.

\subsection{Baseline}
\label{ssec:baseline}

We adopt state-of-the-art CRNN as baseline. The input is Fbank feature of 30-second audios. And the output of our system gives binary predictions for each segment with time resolution of 80 ms (4 times of the input frame shift 20 ms).

The CRNN architecture consists of three parts: convolutional neural network (CNN), recurrent neural network (RNN) and fully-connected layer. The architecture of our CRNN is similar to that in \cite{Kao}, and it is shown in Figure 2.

The CNN part contains four convolutional layers, and each layer is followed by batch normalization \cite{BN}, ReLU activation unit and dropout layer \cite{dropout}. We add two residual connections \cite{resnet} to improve the performance of CNN. Max-pooling layers (on both time axis and frequency axis) are used to maintain the most important information on each feature map. At the end of CNN, the extracted features over different convolutional channels are stacked along the frequency axis.

The RNN part is a bi-directional gated recurrent unit (bi-GRU) layer. Compared with uni-directional GRU, bi-GRU can extract temporal structures of sound events better. We add the outputs of forward GRU and backward GRU to get final outputs of bi-GRU. The size of the output of bi-GRU is (375, \emph{U}), where \emph{U} is the number of GRU units.

After the bi-GRU, a single fully-connected layer with sigmoid activation is used to give classification result for each segment (4 frames). The output denotes the presence probabilities of the target event in each segment.

In order to determine the presence of an event, a binary prediction is given for each segment with a constant threshold of 0.5. These predictions are post-processed with a median filter of length 240 ms. Since at most one event would occur in a 30-s audio, we select the longest continuous sequence of positive predictions to get the onset and offset of target events.

\begin{figure}[tb]
\centerline{\includegraphics[width=0.95\linewidth]{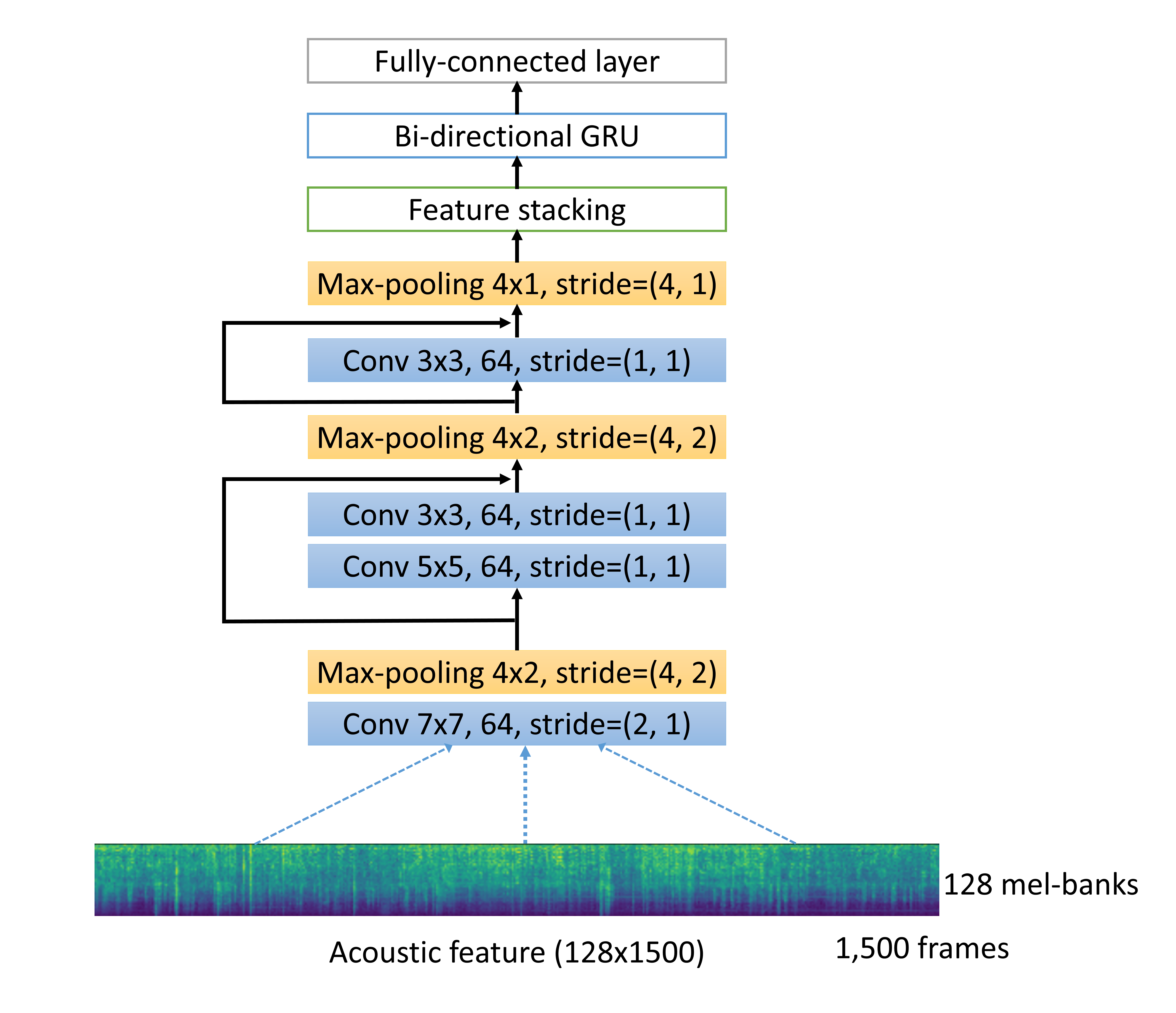}}
\caption{The architecture of CRNN. The first and second dimensions of convolutional kernels and strides represent the time axis and frequency axis respectively.}
\label{fig:fig2}
\end{figure}

\subsection{Learning when to listen}
\label{ssec:time_attention}
As shown in Figure 1, we add a temporal attention model at the end of CNN to enable our system to learn when to listen. This attention model was proposed to ignore irrelevant sounds and focus more on important segments. Unlike the attention model in audio classification \cite{Kong} that only focuses on positive segments (including events), our temporal attention pays more attention to both positive segments and hard negative segments (only backgrounds, but easily misclassified as events) because they should be further differentiated.

The output of CNN will pass through a fully-connected layer with $N_t$ hidden units, followed by an activation unit (sigmoid, ReLU, or softmax). Then a global max-pooling on the frequency axis is used to get one weight for each segment. Those attention weights will be normalized along time axis. In our experiments, this operation of normalization has shown great effectiveness because it takes into account the variation of weight factors along time axis instead of considering only current segment. Then we multiply the temporal attention weights with the output of the fully-connected layer after bi-GRU. A sigmoid function is used to normalize the probabilities to $[0, 1]$. The final output can be computed as follows:

\begin{equation}
  \label{eqn:ta_equation1}
   \hat{a}_t = \max\limits_{n \in \{1, 2, 3, \ldots, N_t\}}{\{\sigma(W_nC_t+b_n)\}},\\
\end{equation}

\begin{equation}
  \label{eqn:ta_equation2}
    a_t = T\frac{\hat{a}_t}{\sum_{t}{\hat{a}_t}},
\end{equation}

\begin{equation}
  \label{eqn:ta_equation3}
   y_t = \frac{1}{1+{\rm exp}(-a_{t}h_{t})},
\end{equation}
where $\sigma(\cdot)$ is an activation function, $C_t$ denotes the output of CNN, $W_n$ and $b_n$ represent the weights and bias for the $n$-th hidden unit respectively, $n \in \{1, 2, 3, \ldots, N_t\}$ and $N_t$ is the number of hidden units in time attention model. $\hat{a}_t$ is the candidate temporal attention weight, $T$ is the total number of segments in an audio, $a_t$ is the normalized temporal attention weight, and $y_t$ is the final output probabilities.

\begin{table*}[thb]
  \centering
  \caption{Performance of proposed models and other methods, in terms of ER and F-score (\%). *** indicates that class-wise results are not given in related paper. We compare the following models: (1) Baseline: our bi-GRU-based CRNN; (2) CRNN+TA: our bi-GRU-based CRNN with temporal attention model; (3) Proposed: our bi-GRU-based CRNN with temporal-frequential attention model; (4) R-CRNN: Region-based CRNN; (5) 1d-CRNN: DCASE 1st place model; (6) CRNN: DCASE 2nd place model.}
  \begin{tabular}{c|cccc|cccc}
    \hline
     {\multirow{2}*{Model}} & \multicolumn{4}{c|}{ Development Dataset} &  \multicolumn{4}{c}{ Evaluation Dataset} \\

    \cline{2-9}
     ~ & babycry & glassbreak & gunshot & average &	 babycry & glassbreak & gunshot & average\\
    \hline

    Baseline & 0.14$|$92.6 & 0.04$|$98.0 & 0.19$|$89.6 & 0.12$|$93.4 & 0.31$|$83.4 & 0.08$|$95.9 & 0.26$|$85.5  & 0.22$|$88.3\\

    CRNN+TA & 0.14$|$92.8 & 0.03$|$98.4 & 0.17$|$90.9 & 0.11$|$94.0 & 0.25$|$87.4 & 0.05$|$97.4 & 0.18$|$90.6  & 0.16$|$91.8\\

    \textbf{Proposed} & {0.10$|$95.1 }& {0.01$|$99.4} & {0.16$|$91.5} & {0.09$|$95.3} & {0.18$|$91.3} &\textbf{ 0.04$|$98.2} & \textbf{0.17$|$90.8}  & \textbf{0.13$|$93.4}\\

    \hline
    R-CRNN \cite{Kao}& 0.09$|$\;***  & 0.04$|$\;***  & \textbf{0.14$|$\;***}  & 0.09$|$95.5 & ****** & ****** & ****** & 0.23$|$87.9 \\

    1d-CRNN \cite{Lim}& \textbf{0.05$|$97.6} & \textbf{0.01$|$99.6} & 0.16$|$91.6 & \textbf{0.07$|$96.3} & \textbf{0.15$|$92.2} & 0.05$|$97.6 & 0.19$|$89.6  & 0.13$|$93.1\\

    CRNN \cite{Cakir}& ****** & ****** & ****** & 0.14$|$92.9 & 0.18$|$90.8 & 0.10$|$94.7 & 0.23$|$87.4 & 0.17$|$91.0\\


    \hline
    \end{tabular}
  \label{table5}
\end{table*}

\subsection{Learning where to listen}
\label{ssec:freq_attention}

Apart from temporal attention model, we proposed a frequential attention model. As we all know, various sound events may have different spectral characteristics. So we assume that we should treat those frequency components differently based on the characteristic of each frame.

The structure of frequential attention model is similar to temporal attention model. The input Fbank feature will go through a fully-connected layer with $N_f$ hidden units, followed by an activation function (sigmoid, ReLU, or softmax). Here, $N_f$ is set to 128 to correspond with the number of mel-filters. Then it is normalized along the frequency axis to get frequential attention weights. Finally, an element-wise multiplication is adopted between the frequential attention weights and input Fbank feature before the feature is fed into CRNN architecture. The weighted feature is computed as follows:

\begin{equation}
  \label{eqn:fa_equation1}
   \hat{M}_{n, t} = {\sigma(V_nF_t+c_n)},\\
\end{equation}

\begin{equation}
  \label{eqn:fa_equation2}
    {M}_{n, t} = N_f\frac{\hat{M}_{n, t}}{\sum_{n}{\hat{M}_{n, t}}},
\end{equation}

\begin{equation}
  \label{eqn:fa_equation3}
    \tilde{F}_t = M_t \otimes F_t,
\end{equation}
where $\sigma(\cdot)$ is an activation function, $F_t$ is the input acoustic feature, $V_n$ and $c_n$ represent the weights and bias for the $n$-th hidden unit respectively. $\hat{M}_{n, t}$ is the candidate frequential attention weight, ${M}_{n, t}$ is the normalized frequential attention weight, $\otimes$ represents element-wise multiplication and $\tilde{F}_t$ is the weighted feature.

\section{Experiments}
\label{sec:experiments}
\subsection{Dataset}
\label{ssec:dataset}
We demonstrate proposed model on DCASE 2017 Challenge task 2 \cite{task}. The task
dataset consists of isolated sound events for each target class and recordings of everyday
acoustic scenes to serve as background \cite{dataset}. There are three target event
classes: babycry, glassbreak and gunshot.
A synthesizer for creating mixtures at
different event-to-background ratios is also provided. The dataset is comprised of development dataset and evaluation dataset. The development dataset also
consists of two parts: train subset and test subset. Participants are allowed to
use any combination of the provided data for training, and evaluate their models
on the test subset of development dataset. Ranking of submitted systems is based on their performance on evaluation dataset. Detailed information about this task and dataset is available in \cite{task}\cite{dataset}.

We use the synthesizer to generate 3000 mixtures
for each class. The event-to-background ratios are -6, 0, 6dB, and the event presence
probability is set to 0.9 (default value: 0.5) in order to gain more positive samples and mitigate the problem of data imbalance.
We use the development test subset to optimize our model and finally evaluate it on the evaluation dataset.

\begin{figure*}[tb]
 \centering
 \subfigure[Visualization of temporal attention weights]{\includegraphics[width=0.48\linewidth]{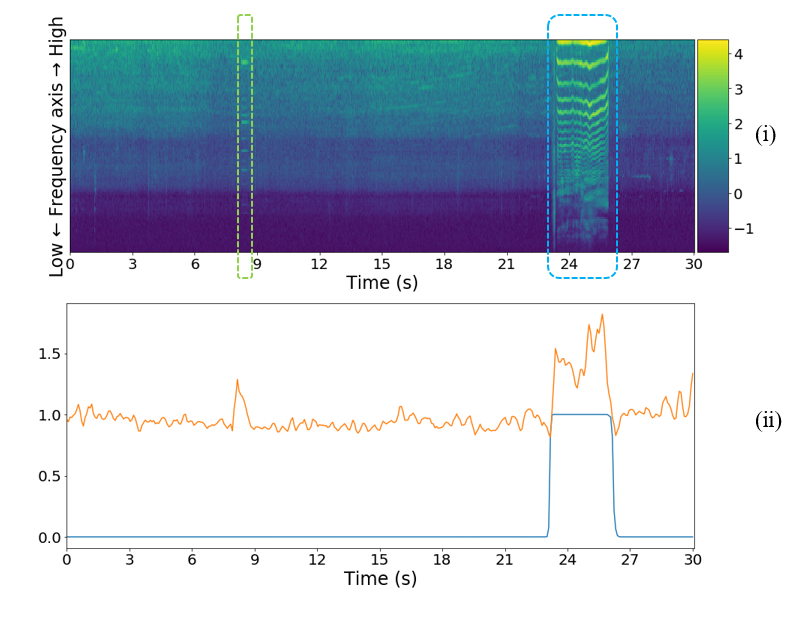}}
 \subfigure[Visualization of frequential attention weights]{\includegraphics[width=0.48\linewidth]{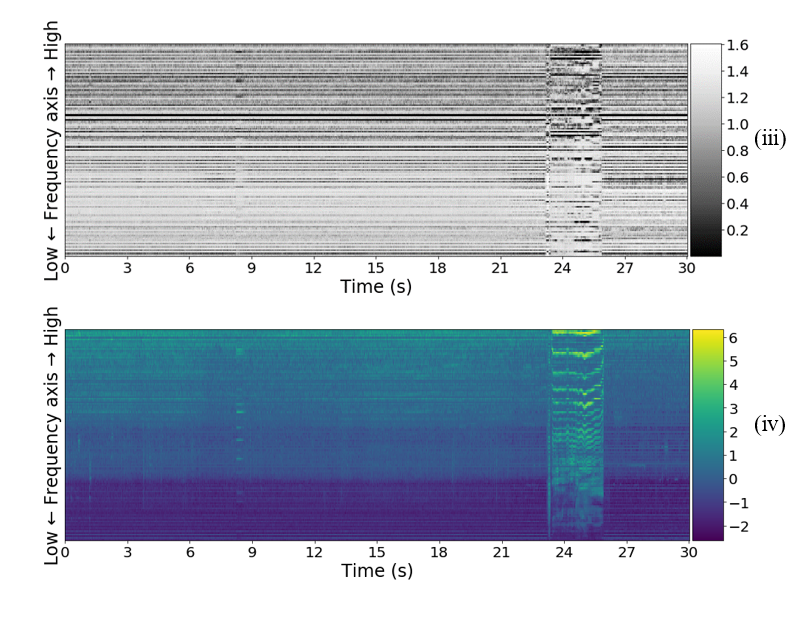}}
\caption{Visualization of attention models.}
\label{fig:res}
\end{figure*}

\subsection{Experiment setup}
\label{ssec:setup}

Our model is trained using Adam \cite{Adam} with learning rate 0.001. Due to data imbalance, we use weighted cross-entropy loss function to reduce deletion error. 
The loss function is computed as follows:
\begin{equation}
  \label{eqn:loss}
   Loss = -\frac{\sum{{\rm w}\hat{y}_t \log(y_t)+(1-\hat{y}_t)\log(1-y_t)}}{N}
\end{equation}
where $y_t$ is the output score of each segment, $\hat{y}_t$ is ground-truth label, and ${\rm w}$ is the loss weight for positive samples. In our experiments, the value of ${\rm w}$ equals to 10.

In order to accelerate training, we adopt pre-training strategy. We firstly train the baseline CRNN for 10 epoches and then use the pre-trained CRNN to initialize the weights during the training of proposed model. The training is stopped after 200 epoches. The batch size is 64. The number of hidden layer unit in temporal attention model $N_t$ is 32. The number of GRU units $U$ is 32.

Because our work is a 0/1 classification system, we use sigmoid and ReLU activation in attention models. According to experimental results, our system can achieve the best performance with ReLU activation in temporal attention model and sigmoid activation in frequential attention model.

\subsection{Metrics}
\label{ssec:Metrics}
We evaluate our method based on two kinds of event-based metrics: event-based error rate (ER) and event-based F-score. Both metrics are computed as defined in \cite{sed_eval}, using a collar of 500 ms and considering only the event onset. If the output accurately predicts the presence of target event and its onset, we denote it as correct detection. The onset detection is considered accurate only when it is predicted within the range of 500 ms of the actual onset time. ER is the sum of deletion error and insertion error, and F-score is the harmonic average of precision and recall. We compute these metrics using sed\_eval toolbox \cite{sed_eval} provided by DCASE organizer.

\section{Results}
\label{sec:result}
\subsection{Experimental results}
\label{ssec:result}
The performances of proposed models and other methods, in terms of ER and F-score, are shown in Table 1.
Results show that temporal attention model can improve the performance of bi-GRU based CRNN baseline, and frequential attention model can make further improvement.
Compared with baseline, proposed method can improve the performance of all classes on both
development dataset and evaluation dataset.

Compared with other state-of-the-art methods, the performance of our model is also competitive. Note that both of the top 2 teams adopt ensemble method. Lim et al. \cite{Lim} combined the output probabilities of more than four models with different time steps and different data mixtures to make final decision. Cakir et al. \cite{Cakir} utilized the ensemble of seven architectures. We can achieve comparable results on development dataset without any model ensemble. Moreover, the average ER only increases slightly from 0.09 to 0.13 on evaluation dataset. We believe that our proposed model has a better capability of generalization. Proposed model achieves the lowest average ER (0.13) and the highest average F-score (93.4\%) on evaluation dataset, outperforming all other methods.

\subsection{Visualization of attention models}
\label{ssec:visualization}
In order to know more about our attention models, we visualize the weights of both temporal attention model and frequential attention model. Presented in Figure 3 is a good example of what our proposed temporal-frequential attention model has actually learnt. Figure 3 (a) and (b) are visualization of temporal attention weights and frequential attention weights respectively.

In Figure 3, (\romannumeral1) is the mel-spectrogram of an audio in the evaluation dataset. In this audio, babycry occurs from 23.13s to 26.16s with ``bus" background. There is a ``beep" sound at around 9-th second. In (\romannumeral2), the blue line denotes the output probability and the orange line denotes the temporal attention weights. We can notice that the weight value is bigger when ``beep" and ``babycry" occur, which conforms with our previous assumption that temporal attention model gives more attention to positive segments and hard negative segments. (\romannumeral3) is the visualization of frequential attention weights and (\romannumeral4) is the spectrogram of weighted feature. We can find that the value of frequential attention weight is bigger in low-frequency area, which means that our frequential attention pays less attention to high frequency components. This can be considered as a low-band filter and frequential attention model can ignore some high-frequency noise.

\section{Conclusion}
\label{sec:conclusion}
In this paper, we proposed a temporal-frequential attention model for sound event detection. Proposed model is tested on DCASE 2017 task 2. Our system can achieve the best performance on DCASE evaluation dataset even without model ensemble. In addition to sound event detection, our temporal-frequential attention model can be applied in speaker verification, speech recognition, audio tagging in the future for further research.

\bibliographystyle{IEEEbib}
\bibliography{strings,refs}
\end{document}